\documentclass[twocolumn,superscriptaddress,amsmath,amssymb,aps,prl,floatfix]{revtex4-2}
\usepackage[usenames,dvipsnames]{color}
\usepackage{xcolor}
\usepackage{multibib} 

\usepackage{comment}
\usepackage{cancel}
\usepackage{ulem}
\usepackage{physics}
\usepackage{graphicx}
\usepackage{dcolumn}
\usepackage{multirow}
\usepackage{bm}
\usepackage{gensymb}
\usepackage{tensor}
\usepackage[percent]{overpic}
\usepackage{tikz}
\usepackage{pgfplots}
\pgfplotsset{compat=1.15}
\usepackage{mathrsfs}
\usetikzlibrary{arrows}
\definecolor{qqttcc}{rgb}{0,0.2,0.8}
\definecolor{ccqqqq}{rgb}{0.8,0,0}
\definecolor{uuuuuu}{rgb}{0.26666666666666666,0.26666666666666666,0.26666666666666666}

\usepackage[breaklinks=true,colorlinks,citecolor=blue,linkcolor=blue,urlcolor=blue]{hyperref}
\definecolor{1}{HTML}{386CB0}
\definecolor{2}{HTML}{2CA02C} 

\begin{document}

\title{Twist-angle transferable continuum model and second flat Chern band in twisted MoTe$_2$ and WSe$_2$}

\author{Xiao-Wei Zhang}
\affiliation{Department of Materials Science and Engineering, University of Washington, Seattle, Washington 98195, USA}
\author{Kaijie Yang}
\affiliation{Department of Materials Science and Engineering, University of Washington, Seattle, Washington 98195, USA}
\author{Chong Wang}
\affiliation{Department of Materials Science and Engineering, University of Washington, Seattle, Washington 98195, USA}
\author{Xiaoyu Liu}
\affiliation{Department of Materials Science and Engineering, University of Washington, Seattle, Washington 98195, USA}
\author{Ting Cao}
\email{tingcao@uw.edu}
\affiliation{Department of Materials Science and Engineering, University of Washington, Seattle, Washington 98195, USA}
\author{Di Xiao}
\email{dixiao@uw.edu}
\affiliation{Department of Materials Science and Engineering, University of Washington, Seattle, Washington 98195, USA}
\affiliation{Department of Physics, University of Washington, Seattle, Washington 98195, USA}
\affiliation{Pacific Northwest National Laboratory, Richland, Washington 99354, USA}

\begin{abstract}
We develop a twist-angle transferable continuum model for twisted transition metal dichalcogenide (tTMD) homobilayers, using tMoTe$_2$ and tWSe$_2$ as examples. All model parameters are extracted from density functional theory (DFT) calculations at a single twist angle (3.89°) and monolayer data.
Our model captures both lattice relaxation effects and the long-range behavior of piezoelectric and ferroelectric potentials.
Leveraging lattice relaxations obtained via machine learning force fields (MLFFs), the model can be efficiently transferred to other twist angles without requiring additional DFT calculations. It accurately reproduces the DFT band dispersions and quantum geometries across a wide range of twist angles. 
Furthermore, our model reveals that a second flat Chern band arises near 2$^\circ$ when the interlayer potential difference becomes comparable to the interlayer tunneling.
This continuum model provides a clear understanding and starting point for engineering novel electronic phases in moir\'e TMDs through twist angles and lattice relaxations.
\end{abstract}
\maketitle
In recent years, substantial experimental progress has been made in observing novel electronic phases due to the interplay between topology and correlations in $R$-type moir\'e TMD homobilayers~\cite{cai2023signatures,zeng2023thermodynamic,park2023observation,xu2023observation,foutty2024mapping,kang2024evidence,nuckolls2024microscopic,ji2024local,anderson2024trion,redekop2024direct,park2025ferromagnetism,wang2025hidden,thompson2024visualizing,xia2025superconductivity,guo2025superconductivity}, including zero-field fractional Chern insulators~\cite{cai2023signatures,zeng2023thermodynamic,park2023observation,xu2023observation} and superconductivity~\cite{xia2025superconductivity,guo2025superconductivity}.
Theoretically, continuum models are widely used as a single-particle starting point for studying these phases. However, constructing such models remains challenging due to significant lattice relaxation effects. Unlike twisted bilayer graphene, $R$-type tTMDs permit moiré ferroelectricity and piezoelectricity, arising from local inversion symmetry breaking. Prior DFT studies have demonstrated that the twist-angle dependent competition between these two polarizations plays a crucial role in shaping moiré band structures and quantum geometries~\cite{zhang2024polarization,jia2024moire,mao2024transfer,wang2024fractional}.
Notably, multiple flat bands with Chern number $C = 1$ were first predicted near $2^\circ$ in tMoTe$_2$~\cite{zhang2024polarization} and signatures of non-Abelian states were later proposed at half-filling of the second flat band~\cite{ahn2024non,reddy2024non,xu2025multiple,chen2025robust,wang2025higher}. However, the physical origin of the second flat Chern band is still unclear based on the current continuum models.

In developing the continuum model, an initial version was constructed using first-harmonic intralayer potentials and interlayer tunneling, with parameters obtained by fitting DFT band structures at various local stackings~\cite{wu2019topological}. However, this model fails to accurately describe the system due to the absence of moiré-scale lattice relaxation~\cite{wang2024fractional}. Subsequent approaches have involved parameterizing the continuum model by fitting large-scale DFT moir\'e band structures and incorporating higher harmonics to account for relaxations~\cite{wang2024fractional,ahn2024non,mao2024transfer,xu2025multiple}. Nevertheless, such fitting procedures introduce inconsistencies across twist angles due to the vast parameter space involved. Alternatively, some efforts have focused on incorporating additional terms—such as those accounting for corrugations and piezoelectric effects—directly into the model~\cite{magorrian2021multifaceted,tang2021geometric}. Yet, the resulting models are still not accurate due to the incomplete treatment of lattice relaxation effects and long-range ferroelectric potentials.

More recently, a precise parameterization method has been developed, which projects the DFT Hamiltonian directly onto the continuum model basis without the need for fitting~\cite{zhang2024universal}. Although this approach is exact, it remains computationally expensive, as large-scale DFT calculations are required for each twist angle. Therefore, striking a balance between efficiency and accuracy remains challenging.

In this work, we develop a twist-angle transferable continuum model for $R$-type tWSe$_2$ and tMoTe$_2$. The model comprehensively captures lattice relaxation effects and the long-range behavior of piezoelectric and ferroelectric electrostatic potentials, based on spatially varying interlayer distance $d(\bm{r})$ and in-plane atomic displacement field $\bm{u}(\bm{r})$. The model is parameterized at a single twist angle (3.89$^\circ$) by fitting large-scale DFT band structures, supplemented with monolayer data.
With lattice relaxations provided by MLFFs, the model can be efficiently transferred to arbitrary twist angles without requiring additional DFT calculations. Its accuracy is validated by reproducing DFT band dispersions and quantum geometries across the 2$^\circ$–5$^\circ$ range.
By analyzing the twist-angle dependence of the moir\'e potentials generated by the model, we uncover the origin of a second flat Chern band emerging near 2$^\circ$ in both systems. This feature arises because the interlayer potential differences and interlayer tunneling become comparable as $\theta$ varies. The developed continuum model is generalizable to other moir\'e TMD systems and offers a powerful basis for engineering novel electronic phases through twist angle tuning, strain, and pressure.

\begin{figure*}[t]
\centering
\includegraphics[width=0.95\textwidth]{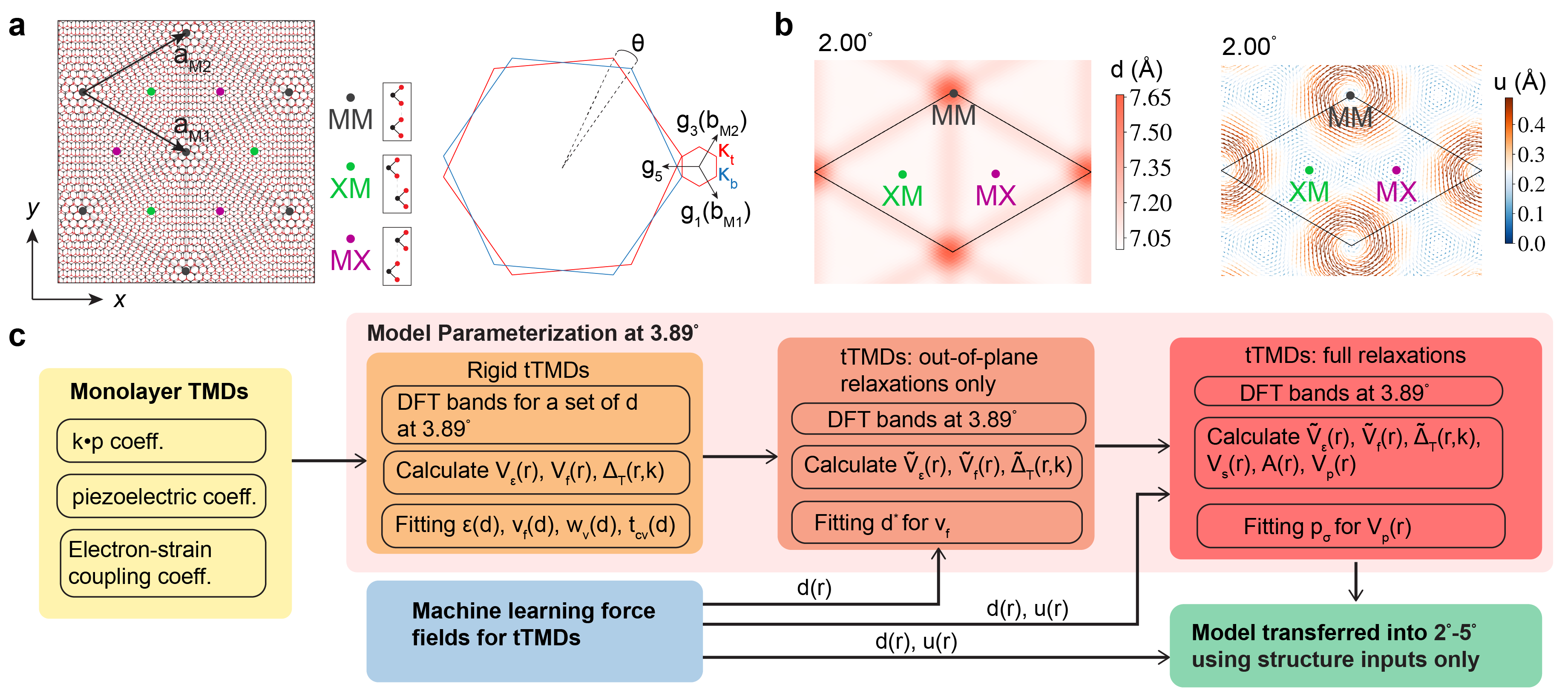}
	\caption{\textbf{Moir\'e superlattice, lattice relaxations, and model parameterization}  \textbf{a}, schematic illustration of the moir\'e superlattice and the moir\'e Brillouin zone. The three high-symmetry stackings—MM, XM, and MX—are indicated. \textbf{b}, interlayer distance (ILD) distribution (left) and top-layer in-plane atomic displacements (right) for 2.00$^\circ$ tMoTe$_2$. Colors represent the magnitude of in-plane displacements, while arrows indicate their directions. \textbf{c}, workflow for model parameterization using DFT calculations and subsequent model predictions incorporating moiré lattice relaxations obtained from MLFFs.
     }
	\label{fig1}
\end{figure*}
We begin by incorporating lattice relaxation and long-range polarization effects into our model.
In $R$-type TMDs, the lack of inversion symmetry in monolayer TMDs allows intralayer piezoelectric polarizations induced by strain fields in each layer~\cite{duerloo2012intrinsic}. Because the two layers have opposite atomic displacements and the same piezoelectric coefficients, the piezoelectric polarizations can cause a vertical potential drop across the bilayer~\cite{zhang2024polarization, enaldiev2020stacking,magorrian2021multifaceted}. 
Moreover, the inversion symmetry breaking at local stackings in $R$-type tTMDs permits moiré ferroelectric polarizations~\cite{wu2021sliding,wang2022interfacial,nuckolls2024microscopic}. These polarization charges generate long-range electrostatic potentials. While prior studies have considered the long-range nature of the piezoelectric potential, the ferroelectric potential has typically been treated as a local stacking-dependent effect~\cite{magorrian2021multifaceted}. Here, we first analyze a rigidly twisted moir\'e superlattice to capture the long-range ferroelectric potential, and then incorporate the effects of lattice relaxation.

\textit{Rigid lattice}.---In the rigid lattice, the top layer is rotated by $\theta$ relative to the bottom layer and lattice relaxations are excluded [Fig.~\ref{fig1}(a)]. As a result, the local shift vector between the two layers is expressed as $\bm{r}_0 = \theta\hat{z}\times \bm{r}$.
In this case, although the piezoelectric polarizations vanish, moir\'e ferroelectric polarizations still exist. 
To fit the bands, we begin with the $\bm{k}\cdot\bm{p}$ continuum model for the spin-up $K$ valley~\cite{bistritzer2011moire,jung2014ab, wu2019topological}:
\begin{widetext}
\begin{equation}
  \mathcal{H}_{K \uparrow} = \left(\begin{array}{c}
    - \frac{(\bm{p} - \hbar\bm{\kappa}
    _b)^2}{2 m^{\ast}} + h_{\text{kin,b}}^{h.o.}+  V_{\varepsilon}(\bm{r})-\frac{V_f(\bm{r})}{2}  \hspace{2em} \hspace*{\fill}
    \Delta_T (\bm{r}, \bm{k})\\
    \Delta_T^{\dagger} (\bm{r}, \bm{k}) \hspace*{\fill} \hspace{2em} -
    \frac{(\bm{p} -\hbar{\bm{\kappa}}_t)^2}
    {2 m^{\ast}} +h_{\text{kin,t}}^{h.o.}+  V_{\varepsilon}(\bm{r})+\frac{V_f(\bm{r})}{2} 
  \end{array}\right). 
  \label{rigid}
  \end{equation}
\end{widetext}
The spin-down $K'$-valley Hamiltonian can be obtained through the time-reversal operation.
Here, $\bm{\kappa}_b$ and $\bm{\kappa}_t$ denote the rotated $K$ point of the bottom and top layer, respectively.
The layer-symmetric potentials $V_{\varepsilon}$ and ferroelectric potential $V_f$ can be expanded using the first-harmonic approximation,
\begin{equation}
    V_{\varepsilon}(\bm{r}) = \varepsilon(d)\sum_{j=1,3,5}^{1s}\text{cos}(\bm{g}_j\cdot\bm{r}),
    \label{Veps}
\end{equation}
and 
\begin{equation}
    V_{f}(\bm{r})=v_f(d,\theta)\sum_{j=1,3,5}^{1s}\text{sin}(\bm{g}_j\cdot\bm{r}).
    \label{Vf}
\end{equation}
Here, $\bm{g}_j$ denotes the moir\'e reciprocal lattice vector (as defined in Fig.~\ref{fig1}(a)) and $d$ is the interlayer distance.  The interlayer tunneling is given by 
\begin{align}
    \Delta_T(\bm{r},\bm{k})=&T_v(\bm{r})-\hbar v_F|\bm{k}-\bm{k}_b|e^{i(\theta_k+\theta/2)}T_{cv}(\bm{r})/E_g \nonumber \\ 
    &-\hbar v_{F}T_{vc}(\bm{r})|\bm{k}-\bm{\kappa}_t|e^{-i(\theta_k-\theta/2)}/E_g.
    \label{DeltaT}
\end{align}
Here, $E_g$ and $v_F$ denote the band gap and Fermi velocity of the monolayer TMD, respectively.
The three interlayer tunneling terms, $T_{v}$, $T_{cv}$, and $T_{vc}$, are all expanded in the first-harmonic form, i.e. 
\begin{equation}
  T_{n_j}(\bm{r})=w_{n_j}(d)(1+e^{-i2\pi j/3}e^{-i\bm{g}_3\cdot\bm{r}}+e^{i2\pi j/3}e^{i\bm{g}_1\cdot\bm{r}}),  
  \label{T}
\end{equation}
where $j=\{0, 1, 2\}$, $n_j=\{v,cv,vc\}$, and $v/c$ denote the valence/conduction band. 

Compared to the previous continuum models~\cite{wu2019topological,magorrian2021multifaceted,tang2021geometric}, our formulation in Eq.~\ref{rigid} introduces three improvements. 
First, we take into account the long-range nature of the ferroelectric potential $V_f$.
Although the ferroelectric potential itself is long-range, it originates from localized charge transfer around the MX/XM regions. By solving the Poisson equation, we derive an analytical expression for the amplitude of $V_f$:
\begin{equation}
    v_f(\theta, d) = \Tilde{v}_f(d)(1 - e^{-g d})/(gd).
    \label{vf}
\end{equation}
Here, $g$ is the modulus of the first-shell moir\'e reciprocal lattice vector.
Including this long-range behavior is crucial for ensuring the model's transferability across different twist angles. 
Second, we incorporate intralayer higher-order kinetic terms, denoted as $h_{kin,b/t}^{h.o.}$. Third, we incorporate high order momentum dependence in the interlayer tunneling by including tunneling processes between the conduction and valence bands. Notably, the second and third terms in Eq.~\ref{DeltaT} correspond to the layer-orbit coupling term introduced in Ref.~\cite{crepel2024chiral}.

To ensure the model's transferability with respect to $\theta$, it is essential to analyze the $\theta$-dependence of various terms. First, the kinetic energy parameters are assumed to be independent of $\theta$. Second, the ferroelectric potential amplitude $v_f$ exhibits an explicit $\theta$-dependence through the function $g(\theta)$ in Eq.~\ref{vf}. Third, since the interlayer tunneling is short-ranged in real space, we assume the corresponding amplitudes $w_v$, $w_{cv}$, and $w_{vc}$ are independent of $\theta$. Lastly, the term $V_\varepsilon$ represents an average potential between the layers arising from local stacking variations, and we likewise assume the parameter $\varepsilon$ is independent of $\theta$. Further details regarding Eq.~\ref{rigid} can be found in the Supplementary Note 3.

\textit{Relaxed lattice}.---
We next incorporate lattice relaxation effects into the continuum model.
Figure~\ref{fig1}(b) illustrates the interlayer distance distribution and in-plane atomic displacements for 2.00$^\circ$ tMoTe$_2$.
Consistent with previous studies~\cite{zhang2024polarization, thompson2024visualizing}, triangular domain formations are observed.
These significant corrugations render the parameters $\varepsilon$, $w_v$, and $w_{cv}$ spatially dependent through substitution with the spatially varying interlayer distance $d(\bm{r})$.
However, for the parameter $\Tilde{v}_f$, the spatial variation of $d(\bm{r})$ complicates the solution of the Poisson equation.
To retain analytical tractability, we introduce an effective interlayer distance, $d^\ast$, obtained as the spatial average of $d(\bm{r})$.
Details regarding the determination of $d^\ast$ are provided in Supplementary Note 2.

The in-plane relaxations introduce three distinct effects. First, the local shift vector $\bm{r}_0$ is modified as follows
\begin{equation}
\bm{r}_0 = \theta \hat{z} \times \bm{r} + \bm{u}_t - \bm{u}_b \label{local},
\end{equation}
where $\bm{u}_t$ and $\bm{u}_b$ denote the atomic displacements of the top and bottom layers, respectively~\cite{balents2019general}. Second, the strain field induces scalar and pseudo-vector potentials, given by
  $V_s(\bm{r}) = \zeta_1 (u_{xx} + u_{yy})$ and $\bm{A}(\bm{r}) = -(\hbar \zeta_2)/(ea)(u_{xx} - u_{yy}, -2u_{xy})$~\cite{cazalilla2014quantum, fang2018electronic}. 
Here, $a$ is the lattice constant of the monolayer TMD, $e$ is the elementary charge, and $\zeta_1$ and $\zeta_2$ are $\bm{k}\cdot\bm{p}$ parameters for a strained monolayer TMD.
Third, the strain gradient induces piezoelectric charge densities in each layer, expressed as $\rho_{\text{piezo}} = \Tilde{e}_{11} [2 \partial_x u_{xy} + \partial_y (u_{xx} - u_{yy})]$, which generate piezoelectric potentials $V_{p,t/b}(\bm{r})$~\cite{enaldiev2020stacking, magorrian2021multifaceted}. 
Similar to $V_f$, the long-range nature of $V_{p,t/b}$ is captured by solving the Poisson equation, with the expressions provided in Supplementary Note 2. To account for screening effects from the TMD layers themselves, we introduce a spatially uniform screening factor $p_\sigma$.
\begin{figure*}[t]
\centering
\includegraphics[width=1.0\textwidth]{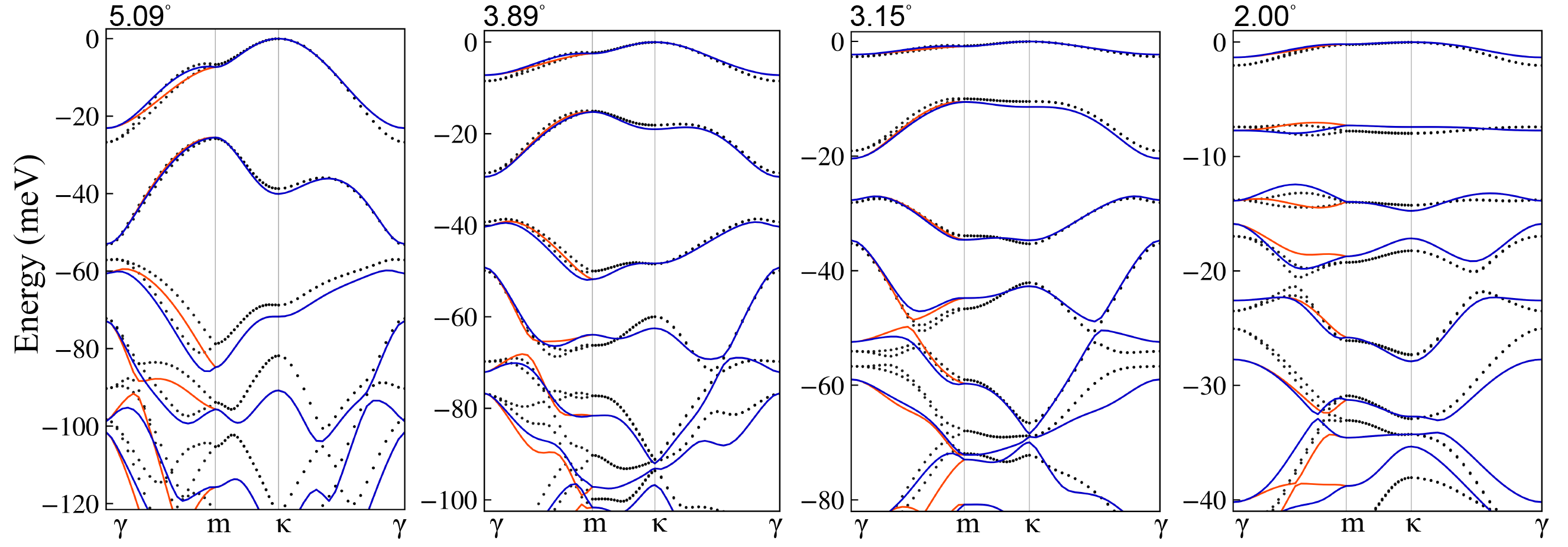}
	\caption{\textbf{Moir\'e band structures.} Comparison of band dispersions between the continuum model and DFT for 5.08$^\circ$, 3.89$^\circ$, 3.15$^\circ$, and 2.00$^\circ$ tMoTe$_2$. Black dots correspond to the DFT results, and red (blue) lines denote the spin-up (spin-down) bands at the $K$ ($K'$) valley obtained from the continuum model. All continuum model parameters are fitted at 3.89$^\circ$.
     }
	\label{fig2}
\end{figure*}
\begin{figure*}[t]
\includegraphics[width=1.0\linewidth]{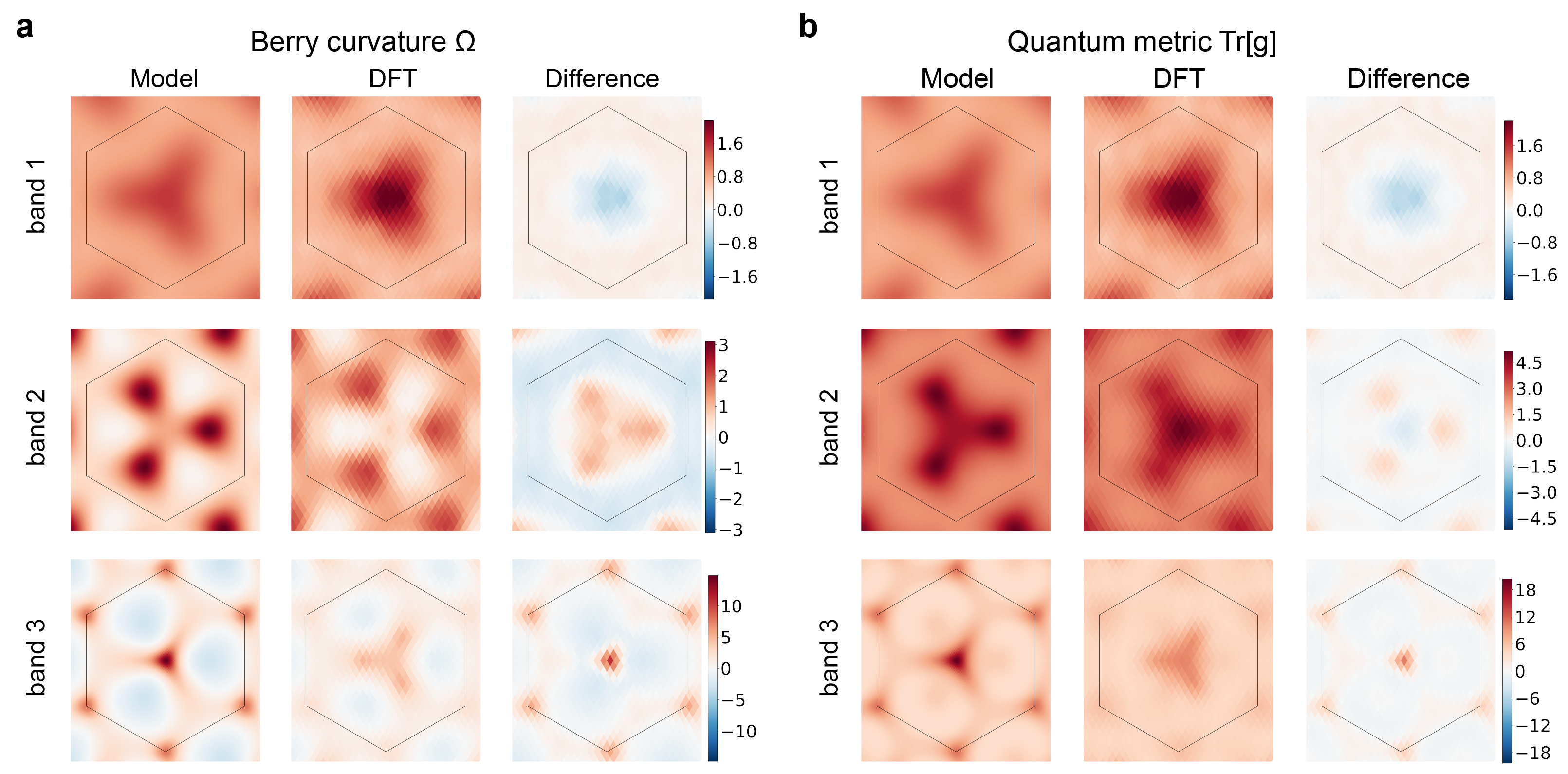}
	\caption{ \textbf{Distributions of Berry curvature and quantum metric of 2.00$^\circ$ tMoTe$_2$.}  \textbf{a}, Berry curvature ($\Omega$) distributions in the first mBZ for the first three $K$-valley moir\'e bands, obtained from the continuum model (left), DFT-based Wannier interpolations (middle), and their differences (right).  
\textbf{b}, similar to \textbf{a}, but shows the distribution of the trace of the quantum metric, Tr($g$), in the mBZ.  $\Omega$ and Tr($g$) are multiplied by the mBZ area and divided by $2\pi$. The black hexagon outlines the first mBZ.
    }
	\label{fig3}
\end{figure*}

\begin{figure*}
\includegraphics[width=1.0\linewidth]{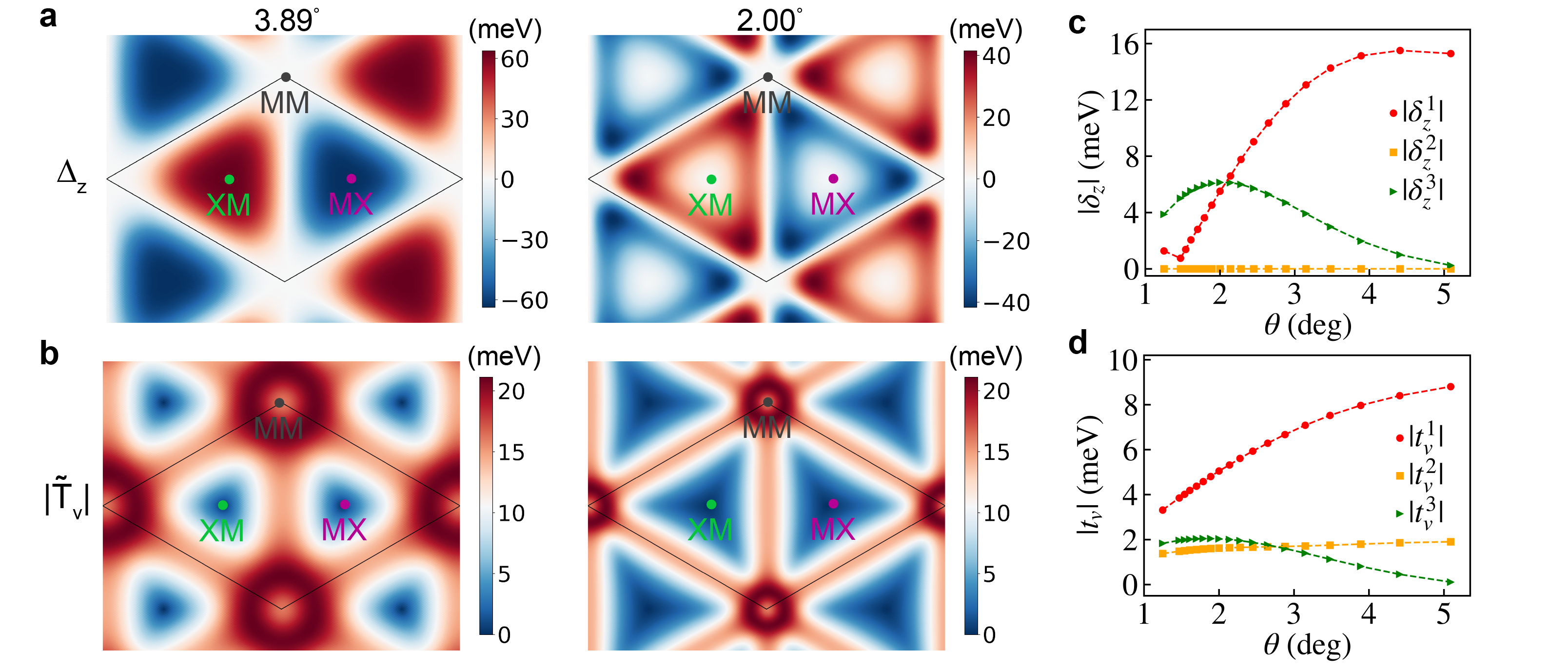}
	\caption{\textbf{Twist-angle dependence of moir\'e potentials in tMoTe$_2$.}  
\textbf{a-b}, real-space distributions of  $\Delta_z(\bm{r})$ and $|\Tilde{T}_v(\bm{r})|$,at 3.89$^\circ$ and 2.00$^\circ$. The black parallelogram outlines a moir\'e unit cell.  
\textbf{c–d}, twist-angle dependence of the modulus of the Fourier components for $\Delta_z(\bm{r})$ and $\Tilde{T}_v(\bm{r})$. 
$\delta_z^i$ and $t_v^i$ denote the $i$th-shell harmonic contribution for $\Delta_z(\bm{r})$ and $\Tilde{T}_v(\bm{r})$, respectively.
       }
	\label{fig4}
\end{figure*}

The total Hamiltonian incorporating lattice relaxation effects can now be written as
\begin{widetext}
\begin{equation}
  \mathcal{\Tilde{H}}_{K \uparrow} = \left(\begin{array}{c}
    - \frac{(\bm{p} - \hbar \bm{\kappa}_b + e\bm{A}_b)^2} {2 m^{\ast}} + h_{\text{kin,b}}^{h.o.}+ \Delta_b(\bm{r})  \hspace{3em} \hspace*{\fill}
    \Tilde{\Delta}_T (\bm{r}, \bm{k})\\
    \Tilde{\Delta}_T^{\dagger} (\bm{r}, \bm{k}) \hspace*{\fill} \hspace{3em} -
    \frac{(\bm{p} - \hbar \bm{\kappa}_t+ e\bm{A}_t)^2
    }{2 m^{\ast}} +h_{\text{kin,t}}^{h.o.}+ \Delta_t(\bm{r})
  \end{array}\right).
  \label{total}
\end{equation}
\end{widetext}
Here, $\Delta_b$ and $\Delta_t$ denote the bottom and top-layer potentials, respectively, i.e. 
\begin{equation}
    \Delta_b(\bm{r}) = \Tilde{V}_{\varepsilon}(\bm{r}) - \Tilde{V}_f(\bm{r})/2 + V_{p,b}(\bm{r})+V_{s,b}(\bm{r}) \label{delta_b}
\end{equation}
and 
\begin{equation}
    \Delta_t(\bm{r}) = \Tilde{V}_{\varepsilon} (\bm{r}) + \Tilde{V}_f (\bm{r})/2 + V_{p,t}(\bm{r}) + V_{s,t}(\bm{r}).
    \label{delta_t}
\end{equation}
Note that the formulations of $\Tilde{V}_{\varepsilon}$, $\Tilde{V}_{f}$, and $\Tilde{\Delta}_T$ slightly differ from their definitions in Eqs.~\ref{Veps}–\ref{DeltaT} due to the modifications of local shift vector as shown in Eq.~\ref{local} (see Supplementary Note 3).
Since all model parameters can be extracted from a single twist angle (in practice, 3.89$^\circ$ is selected for its computational efficiency and experimental relevance) along with monolayer TMD inputs, the continuum model at arbitrary twist angles can be constructed directly from the moir\'e structural information obtained through MLFFs, without requiring additional large-scale DFT calculations.
Equation~\ref{total} represents the central result of this work.

We now briefly outline the parameterization workflow of the continuum model, as illustrated in Fig.~\ref{fig1}(c). First, we compute the $\bm{k} \cdot \bm{p}$ coefficients, piezoelectric coefficient $\Tilde{e}_{11}$, and electron-strain coupling coefficients $\xi_1$ and $\xi_2$ for monolayer TMDs. 
Next, we extract $\varepsilon(d)$, $\Tilde{v}_f(d)$, $w_v(d)$, and $w_{cv}(d)$ in Eqs.~~\ref{Veps}–\ref{T} by fitting DFT moiré bands over a series of interlayer distances at 3.89°.
The effective interlayer distance $d^\ast$ for $\Tilde{V}_f$ is then determined by fitting DFT bands computed with out-of-plane relaxations only. Subsequently, with both in-plane and out-of-plane relaxations included, we determine the screening factor $p_\sigma$ for the piezoelectric potential $V_{p,t/b}$ by further fitting to DFT moiré bands at 3.89$^\circ$. Additional details of the parameterization are provided in Supplementary Note 2, where all the model parameters are tabulated.

\textit{Model validation and transferability}.---Next, we validate the continuum model by comparing its predictions with DFT results. As shown in Fig.~\ref {fig2}, moiré band structures from the model are benchmarked against DFT calculations across a range of twist angles from 2.00° to 5.08°. At 3.89°, where the model parameters are fitted, excellent agreement is observed for at least the top three valence moiré bands in each valley. In addition, the model accurately captures the valley splittings along the $\gamma$–$m$ path for the top bands, primarily due to $T_{cv}$ ($T_{vc}$) and trigonal warping. Notably, the model maintains good agreement with DFT across the full twist angle range. The generalizability of the continuum model is further demonstrated for tWSe$_2$, as shown in Supplementary Note 2.

To further assess the model, we compare its quantum geometries with those obtained from Wannier-interpolated DFT calculations. In Fig.~\ref{fig3}(a) and (b), we show the distributions of Berry curvature $\Omega$ and the trace of the quantum metric Tr($g$) for the first three $K$-valley moir\'e bands at 2.00$^\circ$ for tMoTe$_2$. The model successfully reproduces the momentum dependence of both quantities, as evidenced by the overall small differences shown in the right columns of ~\ref{fig3}(a) and (b). Integration over the moir\'e Brillouin zone yields identical Chern numbers $(+1, +1, +1)$ for both the continuum model and DFT results. Furthermore, the integrated values of Tr($g$) from the model—(1.04, 3.13, 5.26)—closely match those from the DFT—(1.04, 3.09, 5.11)~\cite{wang2025higher}.

\begin{figure*}[t]
\includegraphics[width=1.0\linewidth]{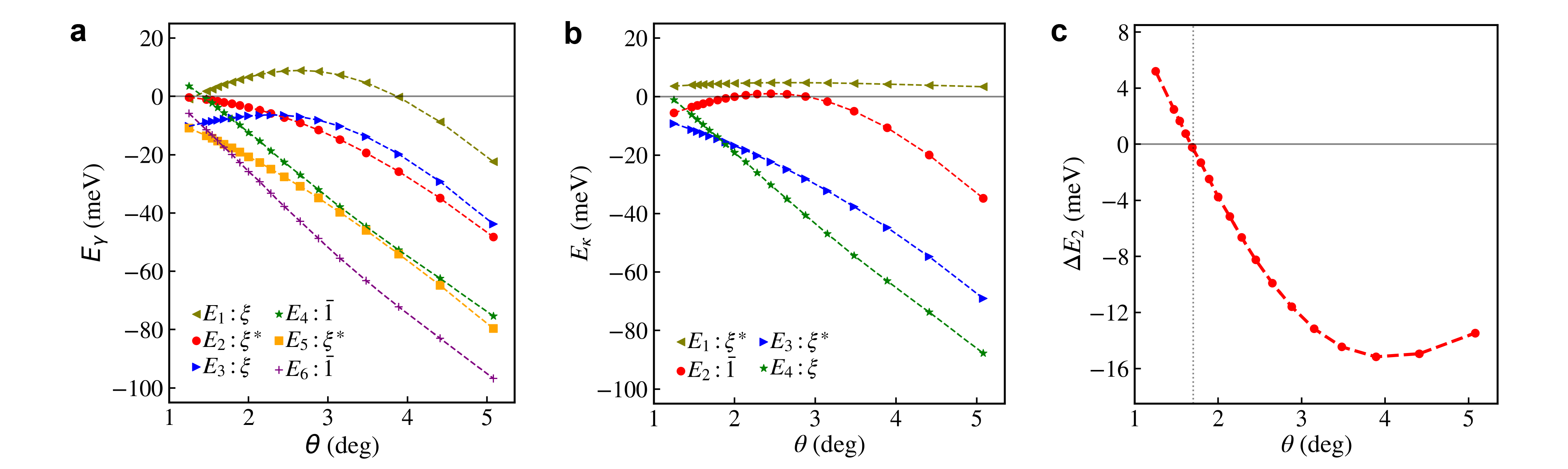}
	\caption{\textbf{Perturbed band energies of tMoTe$_2$.}  
\textbf{a}, twist-angle dependence of the six perturbed eigenvalues at the $\gamma$ point. The eigenvalues are labeled according to their spinful $\hat{C}_3$ eigenvalues: $\xi = e^{i\pi/3}$, $\xi^\ast = e^{-i\pi/3}$, and $\bar{1} = -1$.  
\textbf{b}, twist-angle dependence of the four perturbed eigenvalues at the $\kappa$ point.  
\textbf{c}, twist-angle dependence of the second eigenvalue difference between the $\gamma$ and $\kappa$ points, defined as $\Delta E_2 = E_{\gamma2} - E_{\kappa2}$.
    }
	\label{fig5}
\end{figure*}

\textit{Origin of second flat Chern band}.--- The twist-angle transferability of the continuum model enables us to 
investigate the physical origin of the second flat Chern band emerging around 2.00$^\circ$, a feature first predicted by DFT calculations for tMoTe$_2$ and tWSe$_2$~\cite{zhang2024polarization} (also see Supplementary Note 1). Through a pertubative analysis of the energy and $C_3$ eigenvales at the $\gamma$ and $\kappa$ points, we find that the main reason lies in the twist-angle dependence of $\Tilde{T}_v(\bm{r})$ and the interlayer potential difference $\Delta_z$ which is defined as
\begin{equation}
    \Delta_z(\bm{r}) = \frac{\Delta_b(\bm{r}) - \Delta_t(\bm{r})}{2}.
\end{equation}
Thus, in Fig.~\ref{fig4}(a), we present the spatial distribution of $\Delta_z(\bm{r})$ at 3.89$^\circ$ and 2.00$^\circ$. The primary contributions to $\Delta_z$ come from the ferroelectric potential $\Tilde{V}_f$ and the piezoelectric potential $V_{p,t/b}$, which exhibit significant twist-angle-dependent competition~\cite{zhang2024polarization}, as detailed in Supplementary Note 3. 
This evolution is clearly reflected in the Fourier components of $\Delta_z$, shown in Fig.~\ref{fig4}(c). As $\theta$ decreases, the first harmonic $|\delta_z^1|$ drops rapidly and approaches zero near 1.5$^\circ$, before rising again at smaller angles. This non-monotonic behavior arises from a reversal in the dominant contribution—from $V_{p,t/b}$ to $\Tilde{V}_f$—in the MX/XM regions, resulting in a sign flip of $\Delta_z$. Meanwhile, the third harmonic $|\delta_z^3|$ increases with decreasing $\theta$ above 2.00$^\circ$ and then decreases at smaller angles. In particular, $|\delta_z^2|$ vanishes due to the preserved symmetry $\hat{C}_{2y}T$ of Eq.~\ref{total}, where $T$ denotes time-reversal symmetry and $\hat{C}_{2y}$ is the twofold rotation about the $y$ axis.

Fig.~\ref{fig4}(b) and (d) show the distributions and Fourier components of the interlayer tunneling $\Tilde{T}_v(\bm{r})$, respectively. $\Tilde{T}_v$ is primarily localized around the MM regions, while it vanishes near the MX/XM regions. As $\theta$ decreases from $3.89^\circ$ to $2.00^\circ$, the MM region shrinks, leading to a reduction in the first harmonic $|t_v^1|$. Meanwhile, the higher harmonics remain relatively small compared to $|t_v^1|$. 

Next, we examine how the moir\'e potential terms influence the band energies. Since the topological phase transition and band flattening are mainly governed by $\Delta_z$ and $\Tilde{T}_{v}$,  we neglect the vector potentials $\bm{A}_{u,t/b}$ as well as the interlayer tunneling terms $\Tilde{T}_{cv}$ and $\Tilde{T}_{vc}$, whose effects are discussed in Supplementary Note 4.
Figure~\ref{fig5}(a) shows the $\theta$ dependence of the perturbed $K$-valley eigenvalues at the $\gamma$ point, labeled by their spinful $\hat{C}_3$ eigenvalues. As $\theta$ decreases, we observe an inversion between $E_{\gamma2}$ and $E_{\gamma3}$ near 2.5$^\circ$, while $E_{\gamma1}$ remains the highest in energy down to 1.5$^\circ$. Within the range of 1.5$^\circ$ to 2.5$^\circ$, $E_{\gamma1}$ and $E_{\gamma2}$ carry $\hat{C}_3$ eigenvalues of $(\xi, \xi^\ast)$.
Figure~\ref{fig5}(b) presents the $\theta$-dependence of the perturbed eigenvalues at the $\kappa$ point. Here, $E_{\kappa1}$ remains non-inverted with respect to the other eigenvalues as $\theta$ decreases, whereas $E_{\kappa2}$ and $E_{\kappa3}$ undergo an inversion below 1.5$^\circ$. Above 1.5$^\circ$, $E_{\kappa1}$ and $E_{\kappa2}$ carry $\hat{C}_3$ eigenvalues of $(\xi^\ast, \bar{1})$, respectively. 
Further, in the range of 1.6$^\circ$-2.5$^\circ$, the first and second bands carry Chern numbers of $(+1, +1)$, as inferred from their $\hat{C}_3$ eigenvalues. 
These trends are consistent with both the continuum model and DFT results. 

We then define $\Delta E_2 = E_{\gamma2} - E_{\kappa2}$ to measure the flatness of the second band. As shown in Fig.~\ref{fig5}(c), at large angles, $\Delta E_2$ is negative, while at small angles, it becomes positive, crossing zero near $1.7^\circ$. This behavior can be understood by using the following expressions of $E_{\gamma2}$ and $E_{\kappa2}$, 
\begin{align}
     E_{\gamma2} &= E_{\text{kin}} + 2\delta_0^1 + |2t_v^1 + t_v^2|, \nonumber \\ 
     E_{\kappa2} &=E_{\text{kin}} - \delta_0^1 -\sqrt{3} \text{Im}[\delta_z^1],
\end{align}
where $E_{kin}$ denotes the kinetic energy. 
The strong $\theta$ dependence of $\Delta E_2$ is dominated by the competition between $|t_v^1|$ and $|\delta_z^1|$, as both $|\delta_0^1|$ and $|t_v^2|$ exhibit only weak dependence on $\theta$ (see Fig.~\ref{fig4} and Supplementary Note 3). Notably, due to the $\hat{C}_{2y}T$ symmetry, $\delta_z^1$ is purely imaginary, with $\text{Im}[\delta_z^1]$ remaining negative for $\theta$>$1.5^\circ$. At large twist angles, $|\delta_z^1|$>$|t_v^1|$; conversely, at small angles, $|\delta_z^1|$<$|t_v^1|$ (see Fig.~\ref{fig4}). As a result, $\Delta E_2$ crosses zero at an intermediate $\theta$. A similar trend is observed in tWSe$_2$, as illustrated in Supplementary Note 4, where $\Delta E_2$ crosses zero at a slightly larger $\theta$—consistent with DFT results. This resemblance stems from the similar $\theta$-dependent lattice relaxations and moir\'e potentials in tMoTe$_2$ and tWSe$_2$.

In summary, we have developed a twist-angle transferable continuum model that comprehensively incorporates lattice relaxation effects as well as long-range ferroelectric and piezoelectric potentials in tMoTe$_2$ and tWSe$_2$. All model parameters are extracted from DFT band structures at a single twist angle (3.89$^\circ$) and from monolayer TMD inputs.
The model can be efficiently transferred to other twist angles using only moiré structural data obtained from MLFFs, eliminating the need for additional large-scale DFT calculations. Its transferability is demonstrated over the 2$^\circ$–5$^\circ$ range, showing good agreement with DFT results in predicting both band dispersions and quantum geometries.

In previous continuum models, $\Delta_z$ and $\Tilde{T}_v$ have been used to define the Skyrmion lattice~\cite{wu2019topological,zhang2024polarization,morales2024magic,shi2024adiabatic}. However, as illustrated in Fig.~\ref{fig4}, both $\Delta_z$ and $\Tilde{T}_v$ diminish near the MX/XM stacking regions at 2.00$^\circ$, leading to a vanishing Skyrmion exchange field in those areas. 
This suggests that the adiabatic Skyrmion model may fail to capture the electronic behavior near this twist angle.

Using twist-angle-dependent moir\'e potentials, we identify the origin of the second flat Chern band near 2$^\circ$, arising when the interlayer potential difference becomes comparable to the interlayer tunneling. This finding provides a basis for searching for multiple flat Chern bands in other systems. As the model is structure-based, it also opens the possibility of exploring novel band geometric properties by designing specific lattice relaxation patterns~\cite{van2023rotational,pimenta2023pressure,nguyen2024achieving}. The proposed framework is generalizable to other TMDs and can be extended to multilayer supermoir\'e structures. Looking ahead, the continuum model offers a promising basis for engineering novel electronic phases through strain, pressure, and twist angle, etc.

\textbf{Acknowledgements} 

The development of machine-learning enabled methods and advanced codes was supported by the Computational Materials Sciences Program funded by the U.S. Department of Energy, Office of Science, Basic Energy Sciences, Materials Sciences, and Engineering Division, PNNL FWP 83557. XWZ is supported by DOE Award No.~DE-SC0012509.
This research used resources of the National Energy Research Scientific Computing Center, a DOE Office of Science User Facility supported by the Office of Science of the U.S. Department of Energy under Contract No. DE-AC02-05CH11231 using NERSC award BES-ERCAP0032546, BES-ERCAP0033256, and BES-ERCAP0033507.
This work was also facilitated through the use of advanced computational, storage, and networking infrastructure provided by the Hyak supercomputer system and funded by the University of Washington Molecular Engineering Materials Center at the University of Washington (DMR-2308979). 

\textbf{Author contributions}

XWZ, TC, and DX conceived the project. XWZ carried out the construction of the continuum model and DFT calculations with assistance from KY, TC, and DX. CW performed the Wannier function calculations. XWZ, TC, and DX analyzed the results and wrote the paper with input from YK, CW, and XL.

\textbf{Competing interests}

The authors declare no competing interests.

\textbf{Data availability}

The fitted model parameters, Fourier components of the moiré potentials, and the neural network for the MLFFs have been deposited on GitHub. Additional data related to this paper are available from the corresponding authors upon request.

\textit{Corresponding authors:} Ting Cao (tingcao@uw.edu); Di Xiao (dixiao@uw.edu)


%

\end{document}